\documentclass[superscriptaddress]{revtex4}
\usepackage{amsmath}
\usepackage{amssymb}
\usepackage{graphicx}
\usepackage{bm}
\usepackage{url}

\begin{document}
\title{PLUMED: a portable plugin for free-energy calculations with molecular dynamics}

\author{Massimiliano Bonomi}
\affiliation{Computational Science, Department of Chemistry and Applied Biosciences, ETH Zurich, USI Campus, via G. Buffi 13, 6900 Lugano, Switzerland}
\author{Davide Branduardi}
\affiliation{Computational Science, Department of Chemistry and Applied Biosciences, ETH Zurich, USI Campus, via G. Buffi 13, 6900 Lugano, Switzerland}
\author{Giovanni Bussi}
\affiliation{Universit\`a di Modena e Reggio Emilia and INFM-CNR-S3, via Campi 213/A, 41100 Modena, Italy}
\author{Carlo Camilloni}
\affiliation{Department of Physics, University of Milan and INFN sez. Milano, via Celoria 16, Milano 20133, Italy}
\author{Davide Provasi}
\affiliation{Dept. of Structural and Chemical Biology, Mount Sinai School of Medicine, 1425 Madison Avenue, New York, NY 10029-6574, USA}
\author{Paolo Raiteri}
\affiliation{NRI and Dept. of Applied Chemistry, GPO Box U1987, 6845 Perth (WA), Australia}
\author{Davide Donadio}
\affiliation{Department of Chemistry, University of California Davis, One Shields Avenue, Davis, California 95616, USA}
\author{Fabrizio Marinelli}
\affiliation{International School for Advanced Studies (SISSA), Via Beirut 2-4, 34014 Trieste, Italy}
\author{Fabio Pietrucci}
\affiliation{International School for Advanced Studies (SISSA), Via Beirut 2-4, 34014 Trieste, Italy}
\author{Ricardo A. Broglia}
\affiliation{Department of Physics, University of Milan and INFN sez. Milano, via Celoria 16, Milano 20133, Italy}
\affiliation{The Niels Bohr Institute, University of Copenhagen, Blegdamsvej 17, DK-2100 Copenhagen, Denmark}
\author{Michele Parrinello}
\affiliation{Computational Science, Department of Chemistry and Applied Biosciences, ETH Zurich, USI Campus, via G. Buffi 13, 6900 Lugano, Switzerland}

\begin{abstract}
Here we present a program aimed at free-energy calculations in molecular systems.
It consists of a series of routines that can be interfaced with the most popular classical 
molecular dynamics (MD) codes through a simple patching procedure. 
This leaves the possibility for the user to exploit many different 
MD engines depending on the system simulated and on the computational resources available.  
Free-energy calculations can be performed as a function of many collective variables, 
with a particular focus on biological problems, and using state-of-the-art methods such as 
metadynamics, umbrella sampling and Jarzynski-equation based steered MD.
The present software, written in ANSI-C language, can be easily interfaced with both 
fortran and C/C++ codes.  
\end{abstract}

\maketitle

\vspace{0.5cm}
{\bf PROGRAM SUMMARY}
\vspace{0.5cm}

\begin{small}
\noindent
{\em Manuscript Title:}   PLUMED: a portable plugin for free-energy calculations with molecular dynamics                                   \\
{\em Authors:} Massimiliano Bonomi, Davide Branduardi, Giovanni Bussi, Carlo Camilloni, Davide Provasi, Paolo Raiteri,  
Davide Donadio, Fabrizio Marinelli, Fabio Pietrucci, Ricardo A. Broglia, Michele Parrinello \\
{\em Program Title:}  PLUMED                                     \\
{\em Journal Reference:}                                      \\
{\em Catalogue identifier:}                                   \\
{\em Licensing provisions:}    Lesser GPL                  \\
{\em Distribution format:} tar.gz \\
{\em Programming language:}        ANSI-C                           \\
{\em Computer:}      Any computer capable to run an executable produced by  GCC compiler         \\
{\em Operating system:} Linux operative system, Unix OS-es                     \\
{\em RAM:} Depending on the number of atoms, the method chosen and the collective variables used                           \\
{\em Number of processors used:}     1 or more according the MD engine used                         \\
{\em Supplementary material:}   test suite, documentation, collection of patches, utilities                              \\
{\em Classification:}                     
  23 Statistical Physics and Thermodynamics \\
{\em External routines/libraries:}        none                          \\
{\em Nature of problem:}  calculation of free-energy surfaces for biological and condensed matter systems\\
{\em Solution method:} implementation of various enhanced sampling techniques \\
{\em Unusual features:}
  PLUMED  is not a stand-alone program  but it must be interfaced with a MD code (such as GROMACS, NAMD,  DLPOLY or SANDER) that
needs to be recompiled. Each interface is provided in a patch form \\ 
  {\em Running time:} Depending on the number of atoms, the method chosen and the collective variables used \\
\end{small}




\section{Introduction}
Free-energy calculations have a fundamental role in the understanding
of many natural phenomena ranging from protein folding up 
to polymorphic transitions in solids.
To this end,  molecular dynamics (MD)
has been extensively used together with a series of algorithms 
aimed at  extending its capabilities far beyond those allowed by straightforward MD.
Among some of the most popular {\it enhanced sampling} techniques are 
umbrella sampling~\cite{torrie-valleau,wham1,wham2},
Jarzynski equation based methods~\cite{jarzynski},
adaptive force bias~\cite{darv-poho01jcp} and metadynamics~\cite{metad}.
\\
At the same time, a multitude of MD codes have been developed through the years 
with focus on different fields of application. Some of these programs are more solid-state oriented, with 
a particular attention to the variety of implemented potentials (DL\_POLY~\cite{dlpoly}).
Others are more specialized in biomolecular systems, with specific potentials developed to that scope
(CHARMM~\cite{CHARMM},GROMACS~\cite{Hess:2008p11450} and NAMD~\cite{NAMD}), or implicit solvent capabilities
(CHARMM, AMBER~\cite{amber} and GROMOS~\cite{gromos1,gromos2}).  
Recently, a lot of effort has also been devoted to efficient parallelization, 
allowing  a linear scaling reduction of computational
time with the number of processors. In this respect, NAMD, DESMOND~\cite{desmond},
GROMACS and pmemd in AMBER are currently among the best performing programs. 
\\
This wealth of codes provides the user a wide range of capabilities, but only few of them offer
interfaces for specific free-energy methods, often with a limited set of collective variables. 
In particular, metadynamics has been implemented separately for some of these programs,
and so far the only one freely distributed is GROMETA~\cite{camilloni_protG} for GROMACS.
This prompted us to develop a plugin compatible with many of the aforementioned codes so as to 
facilitate free-energy calculations with a unified input.

\section{Theoretical background}\label{theory}

\subsection{Free-energy methods}\label{methods}

We consider a system made of $N$ atoms characterized by microscopic coordinates
$\bm{r}\in\mathbb{R}^{3N}$ and a potential energy function $U(\bm{r})$. 
We then introduce a set of $d$ collective variables (CVs), $\bm{s}(\bm{r})$, where $\bm{s}\in\mathbb{R}^{d}$.
These variables are used as order parameters, \emph{i.e.} to distinguish between
macroscopically different configurations.
The Helmholtz free energy as a function of these CVs is defined as:
\begin{equation}
F(\bm{s})=-\frac{1}{\beta}\ln\int d\bm{r}\ e^{-\beta U(\bm{r})}\delta( \bm{s}- \bm{s}(\bm{r}) ) + C,
\end{equation}
where $C$ is an immaterial constant.
The free energy $F(\bm{s})$ contains crucial informations about the thermodynamics of the system,
and allows  the calculation of the ensemble average of any observable that depends on the
CVs $\bm{s}$. Moreover, when the CVs are properly chosen,
the free-energy profile can be used to model the out-of-equilibrium
behavior of the system by means of a stochastic
dynamics~\cite{zwan61pr,zwan+01book,yang+06jcp,rait+08acie}.

The simplest way to obtain the Helmholtz free energy from
an unbiased MD simulation is to evaluate the histogram of the visited configurations
in the CVs space $N(\bm{s})$, so that the estimated free energy $\tilde{F}(\bm{s})$ reads:
\begin{equation}
\label{eq:histogram}
\tilde{F}(\bm{s})=-\frac{1}{\beta}\ln N(\bm{s}).
\end{equation}
However, in presence of rare events this method
is completely impractical, since it would require an enormous 
computational time.
Many methods have been proposed to tackle the rare-event problem and to calculate
free-energy profiles.
Some of them are aimed at enhancing the sampling of
the canonical ensemble, so that the free energy is still estimated
by Eq.~(\ref{eq:histogram}).
This class includes methods such as
simulated tempering~\cite{mari-pari92el},
parallel tempering~\cite{hans97cpl,sugi-okam99cpl},
Hamiltonian replica exchange~\cite{fuku+02jcp} and
solute tempering~\cite{liu+05pnas}.
Here we concentrate on a second class of methods, which are based
on collecting configurations in a biased ensemble and require one to select the
set of CVs prior to the simulation.
The prototype of all these methods is umbrella sampling~\cite{torrie-valleau}, where
the simulation is performed with a fixed additional
bias potential $V(\bm{s})$, and the unbiased free energy is recovered as
\begin{equation}
\label{eq:histogram-torrie-valleau}
\tilde{F}(\bm{s})=-\frac{1}{\beta}\ln N(\bm{s})-V(\bm{s}).
\end{equation}
While Equation~(\ref{eq:histogram-torrie-valleau}) is valid for any choice of
the bias potential $V(\bm{s})$, the efficiency of the sampling and convergence properties are strongly
dependent on $V(\bm{s})$. In particular, if an approximate free-energy estimate $\tilde{F}'(\bm{s})$
is available before the simulation, the choice
$V(\bm{s})=-\tilde{F}'(\bm{s})$ would give an approximately flat histogram, thus helping
in overcoming the free-energy barriers.
However, it is rather difficult to have a reliable free-energy estimate before
the simulation.
Various improvements have been introduced so as to refine  the bias
potential on the fly (see, among others, Refs.~\cite{wang-land01prl,darv-poho01jcp,metad,mars+06jpcb}).
We focus here on metadynamics, in its standard form~\cite{metad,error} or in the recently introduced
well-tempered flavor~\cite{Barducci:2008}.
In particular, we consider the direct version of metadynamics, where the bias is acting
directly on the microscopic coordinates. For an excellent review of several variants
of metadynamics see Ref.~\cite{laio-gerv08rpp}.

In standard metadynamics the bias potential is built during the simulation
as a sum of Gaussian functions centered on the previously visited configurations
in the CVs space.
This manner of biasing the evolution by discouraging the visited configurations
was first introduced in the taboo search \cite{taboo} and, in the context of MD, by the
local elevation method~\cite{willy}. The approach is also closely related to the
Wang and Landau algorithm~\cite{wang-land01prl},
adaptive force bias~\cite{darv-poho01jcp} and
self-healing umbrella sampling~\cite{mars+06jpcb}.
In metadynamics, the bias at time $t$ is written as
an integral on the past
trajectory $\bm{r}(t)$:
\begin{equation}
V(\bm{s},t)= \int_0^t\ dt'\omega\exp\left(-
\sum_{i=1}^{d} \frac{(s_i(\bm{r})-s_i(\bm{r}(t'))^2}{2\sigma_i^2}
\right).
\end{equation}
Here $\sigma_i$ is the Gaussian width corresponding to the $i$-th CV
and represents the resolution for that CV, and $\omega$ is the rate
at which the bias grows.
As it was shown empirically~\cite{error} and analytically~\cite{bussi_noneq} for model Langevin dynamics,
in the long run the bias will converge to the negative of the free energy and then oscillate around that profile.
As a consequence, the final histogram will be approximately flat in the CVs space, allowing for an uniform exploration
in spite of the free-energy barriers.
We also observe that the bias performs a work on the system, which needs to be dissipated.
Usually during a metadynamics simulation the thermostat keeps the system in thermal
equilibrium, unless the growth rate of the bias is too large.

Metadynamics has been historically used in two different and complementary manners.
It has been used to escape free-energy minima (see \emph{e.g.} ~Ref.~\cite{ogan+05nature}).
\emph{i.e.} to find a reasonable saddle point out of a local minimum. In this case, metadynamics
should be stopped as soon as the system exits from the minimum and starts exploring
a new region of space. In other applications, it has been used to exhaustively explore the
CV space and reconstruct the free energy.
In the examples presented in this paper, we focus more on this latter application.
Its main advantage over umbrella-sampling technique is that it inherently
explores the region of low free energy first. Here the
simulation should be stopped when the
motion of the CVs becomes diffusive in this region of interest,
and the bias itself can be used as an estimate of the underlying
free energy:
\begin{equation}
\tilde{F}(\bm{s},t)=-V(\bm{s},t)
\end{equation}
(note that if $\bm{s}$ does not include all relevant order parameters the bias
may not converge in a reasonable simulation time).
The free-energy estimate at time $t$, $\tilde{F}(\bm{s},t)$,
is indeed an unbiased estimator of the exact
free energy $F(\bm{s})$~\cite{bussi_noneq}.
However, $\tilde{F}(\bm{s},t)$ fluctuates around $F(\bm{s})$ with an amplitude
which depends on both the diffusion coefficient in the CV
space and on the metadynamics parameters $\omega$ and $\sigma$, and
a more accurate calculation can be performed decreasing $\omega$.
Clearly, a smaller $\omega$ means that more time is required to reconstruct the free-energy landscape, therefore a
compromise needs to be found between speed and accuracy.
One can also exploit the fact that $\tilde{F}(\bm{s},t)$ is an unbiased estimate
of $F(\bm{s})$ at all times, and take the time
average of all the profiles as done, for instance, in Ref.~\cite{mich+04prl}.
However,
as the simulation continues,
configurations of higher and higher free energy are explored and, in order
to take the average it is necessary to force the system to remain inside
the region by a suitable restraining potential.

An alternative approach is the recently introduced well-tempered metadynamics~\cite{Barducci:2008}.
Well-tempered metadynamics
is a variant of the method that solves the problem of the fluctuations in a different way,
and is more suitable for performing free-energy calculations in several dimensions since
it allows avoiding the complication of restraining the dynamics inside a region.
In the well-tempered algorithm, the rate at which the bias is grown is decreased
during the simulation proportional to $e^{-V(\bm{s},t)/\Delta T}$,
where $\Delta T$ is a characteristic energy:
\begin{equation}
V(\bm{s},t)= \int_0^t\ dt'\omega e^{-V(\bm{s}(\bm{r}(t')),t')/\Delta T}\exp\left(-
\sum_{i=1}^{d} \frac{(s_i(\bm{r})-s_i(\bm{r}(t'))^2}{2\sigma_i^2}
\right).
\end{equation}
Over long time, it can be shown that the bias converges to a fraction of the exact free energy:
\begin{equation}
\lim_{t\rightarrow\infty} V(\bm{s},t)= -\frac{\Delta T}{T+\Delta T} F(\bm{s}).
\end{equation}
Conversely, it is possible to estimate the free energy as:
\begin{equation}
\tilde{F}(\bm{s},t)= -\frac{\Delta T+T}{\Delta T} V(\bm{s},t).
\end{equation}
This estimator does not suffer of the fluctuation problem of standard metadynamics.
Moreover at variance with standard metadynamics, the exploration for large times will not
be uniform in the CV space but instead it will satisfy the probability distribution:
\begin{equation}
P(\bm{s})\propto e^{-F(\bm{s})/(T+\Delta T)}.
\end{equation}
Thus, the CVs will be sampled at a finite but arbitrarily high temperature $T+\Delta T$.
This is a rather important feature of well-tempered metadynamics, especially
for $d>1$, since it allows to focus the exploration on the low free-energy regions,
such as the main minima and the saddle points.
The explored free-energy range can be varied by tuning $\Delta T$,
and standard metadynamics is recovered for $\Delta T\rightarrow\infty$.

In practical implementations, the bias is updated with a finite
temporal stride $\tau_G$, so that:
\begin{equation}
V(\bm{s},t)= \sum_{t'=0,\tau_G,2\tau_G,\dots}^{t'<t} W e^{-V(\bm{s}(\bm{r}(t')),t')/\Delta T} \exp\left(-
\sum_{i=1}^{d} \frac{(s_i(\bm{r})-s_i(\bm{r}(t'))^2}{2\sigma_i^2}
\right),
\end{equation}
where $W=\tau_G \omega$ is the height of a single Gaussian.

\subsection{Metadynamics implementation} 

A metadynamics implementation should perform two basic tasks:
(a) keep track of the visited configurations in the CV space or, equivalently, of the shape of the bias potential and
(b) add the proper forces to the microscopic dynamics.

The first task is accomplished by maintaining a list of the Gaussians which have been added to the bias.
This list is dynamic and grows during the simulation. The list is also stored in a file
that can be used to restart a simulation, and to plot the bias with an external utility.

The second task requires evaluation of the bias forces, \emph{i.e.} the derivatives of the bias potential with
respect to the microscopic coordinates $\bm{r}$.
This derivative is calculated using the chain rule:
\begin{equation}
\label{eq:chainrule}
\frac{\partial V(\bm{s},t)}{\partial \bm{r}_i}=\sum_{j=1}^{d}\frac{\partial V(\bm{s},t)}{\partial
s_j}\frac{\partial s_j(\bm{r})}{
\partial \bm{r}_i}.
\end{equation}
The first part of the derivative is simply a sum of analytical derivatives of Gaussian functions.
This sum gets more and more expensive as the simulation proceeds and the number of Gaussians grows~\cite{babi+08jcp}.
Usually this is not  a problem since, if we exclude the simplest test cases, this effort is incomparably smaller than that
of evaluating the force-field. For specific needs, an implementation based on the storage of the
bias potential on a grid could be faster.
The second part of the derivative in Eq.~(\ref{eq:chainrule}) depends on the specific choice for the CVs.
Thus, for each of the CVs that one wants to use, it is necessary to provide routines which, given
the microscopic coordinates, return the value of the CV and its gradients.
Writing and debugging these routines for a large number of CVs requires a noticeable effort.
However, it is worthwhile pointing out that this effort is the main ingredient of many other free-energy methods,
and thus our plugin has been adapted to perform other kinds of free-energy calculations,
such as umbrella sampling~\cite{wham1,wham2} or Jarzynski equation~\cite{jarzynski} based methods,
as it will be discussed in Section~\ref{usage}.

\subsection{Collective variables}

The implementation of many different
CVs is required to deal with the huge variety of problems of interest and to give a proper description of each. 
Here we describe all the possibilities present in the current package.

\begin{itemize}
\item {\bf Atom position.} 
The absolute position of an atom or a group of atoms.
This CV is implemented with several options that allow the user to restrict the bias to a given
direction, \emph{e.g.} $z$, or to bias the position of the particle as projected onto a 
selected segment or, in analogy with the path CV, to bias the atoms distance from a segment.
This variable is not translationally invariant.

\item {\bf Distance.}
The distance between two atoms or, more generally,
the distance between the centers of mass of two groups of atoms identified as $G_1$ and $G_2$:
\begin{eqnarray}
s_{dist}&=&\left| \frac{\sum_{i\in G_1}  m_i\bm{r}_i }{\sum_{i\in G_1}   m_i}  -
\frac{\sum_{i\in G_2}  m_i \bm{r}_i}{\sum_{i\in G_2}   m_i}
\right| \\
&=&\left| \bm{r}_{G_1} - \bm{r}_{G_2}  \right|, 
\end{eqnarray}
where $m_i$  and  $\bm{r}_i$  are the mass and the position of the $i-th$ atom respectively,
and $\bm{r}_{G_1}$ and $\bm{r}_{G_2}$ are the centers of mass of the two groups.
\item {\bf Angle.} 
The angle defined by three atoms or, more generally,
the angle defined by the centers of mass of three groups of atoms identified as $G_1$, $G_2$ and $G_3$:
\begin{eqnarray}
\bm{r}_c&=&\bm{r}_{G_1}-\bm{r}_{G_2}\\
\bm{r}_b&=&\bm{r}_{G_1}-\bm{r}_{G_3}\\
\bm{r}_a&=&\bm{r}_{G_2}-\bm{r}_{G_3}\\
s_{angle}&=& \cos^{-1}\left(\frac{\bm{r}_a^{2} +\bm{r}_c^{2} -\bm{r}_b^{2} }{ 2 |\bm{r}_a||\bm{r}_c|}\right). 
\end{eqnarray}
\item {\bf Torsion.}
The dihedral angle defined by four atoms or, more generally,
the dihedral angle defined by the centers of mass of four groups of atoms
identified as $G_1$, $G_2$, $G_3$ and $G_4$:
 \begin{eqnarray}
\bm{r}_a&=&\bm{r}_{G_4}-\bm{r}_{G_3}\\
\bm{r}_b&=&\bm{r}_{G_2}-\bm{r}_{G_3}\\
\bm{r}_c&=&\bm{r}_{G_3}-\bm{r}_{G_2}\\
\bm{r}_d&=&\bm{r}_{G_1}-\bm{r}_{G_2}\\
s_{torsion}&=& \cos^{-1}\left(\frac{(\bm{r}_a\times \bm{r}_b) \cdot ( \bm{r}_c \times\bm{r}_d) }
          { \vert  \bm{r}_a\times \bm{r}_b \vert \vert \bm{r}_c \times\bm{r}_d   \vert    }\right). 
\end{eqnarray}
 
\item {\bf Minimum distance.} 
The distance between the two closest atoms pertaining to two different groups $G_1$ and $G_2$,
approximately obtained with the following expression:
\begin{equation}
s_{mindist}=\frac{b}{\ln\ \sum_{i\in G_1} \sum_{j\in G_2} \exp\left( \frac{b}{\vert \bm{r}_i - \bm{r}_j \vert  }    \right)},
\end{equation}
where $b$ is a user-supplied smoothing parameter.

\item {\bf Coordination.} 
The coordination number of one atom, or more atoms, with respect to another atom or group of atoms 
(\emph{e.g.} the coordination of an ion with respect to all the water molecules in the simulation box):
\begin{equation} 
s_{coord}=\sum_{i\in G_1} \sum_{j\in G_2} s_{ij},
\end{equation} 
where
\begin{equation} 
s_{ij} = \left\{ \begin{array}{lr}
1                                                                                                                                                                         & \hspace{0.5cm}  \textrm{if}  \hspace{0.5cm} \vert \bm{r}_i - \bm{r}_j \vert < \delta \\
\frac{1-\left(\frac{\vert \bm{r}_i - \bm{r}_j \vert-\delta}{r_0}\right)^n}{ 1-\left(\frac{\vert \bm{r}_i - \bm{r}_j \vert-\delta}{r_0}\right)^m} & \hspace{0.5cm}  \textrm{if}  \hspace{0.5cm} \vert \bm{r}_i - \bm{r}_j \vert \ge \delta . \\
\end{array}\right.
\end{equation}
The user-supplied parameters $r_0$, $\delta$, $n$ and $m$ allow a great flexibility to fine-tune the decay of the switching function, \emph{e.g.} a more accurate counting of the atoms in the coordination shell.
In general a good guess for these parameters can be achieved by looking at the pair distribution function between the first and the second group 
of atoms. A good starting point is to take $\delta$ as the position of the first peak in the pair distribution function, 
$r_0$ as the full width at half maximum of the peak and $n$ and $m$ to force $s_{ij}\simeq0$ at the first minimum of the pair distribution function. 
However, depending on the system properties, different choices may give better results.

\item {\bf Hydrogen bonds.} The number of intra-protein hydrogen bonds, defined as:
\begin{equation}
s_{hbonds}=\sum_{i\in G_O} \sum_{j\in G_H} f(i,j)  \frac{1-\left(\frac{\vert \bm{r}_i - \bm{r}_j \vert}{r_0}\right)^n}{ 1-\left(\frac{\vert \bm{r}_i - \bm{r}_j \vert}{r_0}\right)^m},
\label{coord}
\end{equation}
where $G_O$ the group of oxygen atoms of the protein,
$G_H$ the group of hydrogen atoms of the protein.
Typically, $r_0=2.5 \ {\rm \AA}$, $n=6$ and $m=10$.
The function $f(i,j)$ selects a particular type of hydrogen bonds,
depending on the user choice: all, $\alpha$--helix pattern, $\beta$--strand
pattern (even or odd).

\item {\bf Interfacial water.} 
This variable is intended to calculate the number of atoms of a certain group $G_0$ that are 
in contact with atoms of both groups $G_1$ and $G_2$ at the same time (\emph{e.g.} the number of waters at the interface of two surfaces):    
\begin{equation}
s_{waterbridge}=   \sum_{i\in G_0}  \left( \sum_{j\in G_1} \frac{1-\left(\frac{\vert \bm{r}_i - \bm{r}_j \vert}{r_0}\right)^n}{ 1-\left(\frac{\vert \bm{r}_i - \bm{r}_j \vert}{r_0}\right)^m}  \right )
\left( \sum_{k\in G_2} \frac{1-\left(\frac{\vert \bm{r}_i - \bm{r}_k \vert}{r_0}\right)^n}{ 1-\left(\frac{\vert \bm{r}_i - \bm{r}_k \vert}{r_0}\right)^m}  \right ).
\end{equation}
More precisely, the variable counts the number of atom triples $(i\in G_0,j\in G_1,k\in G_2)$
with $i$ in contact with both $j$ and $k$
(see also coordination parameters). 

\item {\bf Radius of gyration.} 
The radius of gyration of a group $G$, defined as:
\begin{equation}
s_{gyration}=\sqrt{ \frac{\sum_{i\in G}  m_i \vert \bm{r}_i -\bm{r}_G \vert ^2 }{\sum_{i\in G}   m_i} }.
\end{equation}
Similarly, one can be interested in the trace of the inertia tensor, which can be shown
to be equal to:
\begin{equation}
s_{\textrm{Tr}[I]}=
2
(s_{gyration})^2
{\sum_{i\in G}   m_i}.
\end{equation} 

\item {\bf Dipole moment.} 
The electric dipole of a group of atoms:
\begin{equation}
s_{dipole}=\vert \sum_i^{n}  q_i\bm{r}_i \vert, 
\end{equation}
where $q_i$ is the charge of each atom $i$.

\item {\bf Dihedral correlation.} 
This variable measures the degree of similarity of a list of adjacent dihedral angles. 
It is defined by: 
\begin{equation}
s_{dihecorr}=\sqrt{\sum_{i=2}^{n}\left( 1 + \cos^2\left( \frac{\phi_i-\phi_{i-1}}{2}\right)\right)},
\end{equation}
where $\phi_i$ is the dihedral angle defined  by four atoms. 
For proteins, the variable grows with the content of secondary structure.

\item {\bf Alpha-beta similarity.} 
This variable measures the similarity of dihedral angles with respect to reference values:
\begin{equation}
s_{\alpha-\beta}=\frac{1}{2}{\sum_{i=1}^{n}\left( 1 + \cos\left( \phi_i-\phi_{i}^{ref}\right)\right)},
\end{equation}
where reference dihedrals $\phi_{i}^{ref}$ are given as input.
For proteins, this variable can be use to measure the amount of $\alpha$ or $\beta$ 
secondary structure.

\item {\bf Torsional rmsd.} 
 Root of mean square deviation of selected dihedral angles with respect to a reference configuration:
\begin{equation}
s_{tors rmsd}=\sqrt{     \frac{ \sum_i^{n}( \theta_i - \theta_i^{ref} )^2  }{n }},
\end{equation}  
where $n$ is the number of reference dihedrals.

\item {\bf Path collective variables.} 
Path collective variables are a general approach based on a previous (approximate) knowledge
of the reaction path~\cite{brand07}.  
If one assumes that the transition from A to B can be described
by a set
of CVs $\mathbf{S}(\bm{r})$, which are in general
non-linear vectorial functions of the microscopic variables $\bm{r}$, 
then it is possible to define two 
associated variables. One aims at measuring the 
progress along  a parametric path in CVs space  $\mathbf{S}(l)$ composed 
of $P$ frames:
\begin{equation}
s\left( \bm{r}\right) = \frac{\sum_{l=1}^{P} l\ e^{-\lambda
\parallel \mathbf{S}(\bm{r})-\mathbf{S}(l)\parallel ^{2}}}{%
\sum_{l=1}^{P}e^{-\lambda \parallel \mathbf{S}(\bm{r})-\mathbf{S}%
(l)\parallel ^{2}}},  \label{discrets}
\end{equation}
and the other measures the distance from the closest point along the path:
\begin{equation}
z\left( \bm{r}\right) =-\frac{1}{\lambda }\ln \Big(\sum_{l=1}^{P}e^{-%
\lambda \parallel \mathbf{S}(\bm{r})-\mathbf{S}(l)\parallel ^{2}}\Big ),
\label{discretz}
\end{equation}%
where  $\left\Vert \ldots \right\Vert $ is the metric that defines the
distance between two configurations. 
With such definitions $s\left( \bm{r}\right) $ is a pure number that ranges from 1 to $P$ and $z\left( \bm{r}\right)$ has the dimension of  the chosen distance, squared. The  parameter $\lambda$ is in general chosen  as 
$2.3/(\Delta d)^2$ where $\Delta d$ is the average distance among adjacent frames.

The definition of the $\mathbf{S}(l)$ and of the measure $\left\Vert \ldots \right\Vert $ can be chosen by the user among the following:
\begin{itemize} 
\item {\bf Root mean square displacement in Cartesian coordinates}:  The path  $\mathbf{S}(l)$ 
may be defined as a series of configuration in Cartesian space. Each configuration is made of a subset of atoms of the system   and the distance is calculated as  root mean square displacement (RMSD) after optimal alignment \cite{kearsley}.
\item {\bf RMSD in distances:}
the path  $\mathbf{S}(l)$ must be defined as a set of pair distance among atoms. The RMSD is calculated as a difference of distances:
\begin{equation}
 \left\Vert \mathbf{S}(\bm{r})-\mathbf{S}(l)) \right\Vert = 
\sqrt{ \frac{\sum_{i}^{N_{dist}} ( d_{i}(\bm{r}) -d_{i}(l) )^2    }{N_{dist}}},
\end{equation}  
where $d_{i}$ is the distance between the atoms of the $i$-th couple
 and $N_{dist}$ is the total number of couples considered \cite{havel,torda}.
\item {\bf Contact map distance:} the path $\mathbf{S}(l)$ is defined
by a set of contact maps \cite{Bo.Bra:08}, where each contact is defined as in Eq.~(\ref{coord}).
The distance is defined as:    
\begin{equation}
\Vert \mathbf{S}(\bm{r})-\mathbf{S}(l)\Vert =\sqrt{
\sum_{i}^{N_{cont}}(\mathbf{C}_{i}(\bm{r})-\mathbf{C}_{i}(l))^{2}
},
\end{equation}%
where $\mathbf{C}_{i}(\bm{r})$ is the $i$-th contact for the configuration $\bm{r}$ and $N_{cont}$ is the total number of contacts considered.  
\end{itemize}

The CV $z\left( \bm{r}\right)$, when used with a single reference
frame or contact map, is equivalent to the distance between a configuration
and the reference structure, measured in the chosen metric (squared). Therefore, 
this CV should be used to reproduce the standard RMSD, distance RMSD or CMAP distance.

\end{itemize}

\section{Usage examples}\label{usage}
When a given program is instructed to use PLUMED (see the manual for specific implementations),
a supplementary input file for the free-energy calculation must be provided. \\
During the calculation the main output is a file containing a record of the 
values of CVs. This file is generally called {\verb COLVAR }. It may contain also additional informations depending on the chosen free-energy method.
\\
In the following we illustrate the basic use of the different methods, available for all the MD codes specified in 
section \ref{instruction},
and of the algorithms currently implemented only in the GROMACS version.
Many additional examples of CVs and sampling techniques, such as multiple walkers metadynamics \cite{multiplewalkers}, 
 are contained in the test directory distributed with the code.
 
\subsection{Metadynamics} 
A generic input for metadynamics appears as follows: 
\begin{verbatim}
# switching on metadynamics and Gaussian parameters
HILLS HEIGHT 0.1 W_STRIDE 100 
# well-tempered metadynamics
WELLTEMPERED SIMTEMP 300 BIASFACTOR 10
# instruction for CVs printout 
PRINT W_STRIDE 50 
# a simple CV: the distance between atoms
# or  group of atoms (in this case between atom 13 and atoms group <g1>)
DISTANCE LIST 13 <g1> SIGMA 0.35
g1->
17 20 22 30
g1<-
# wall potential
UWALL CV 1 LIMIT  15.0 KAPPA 100.0 EXP 4.0 EPS 1.0 OFF 0.0
# end of the input
ENDMETA
\end{verbatim}

Three kinds of keyword may exist: the \emph{directive} (needed keyword to be placed in the first position 
along a line that specifies the intent of the following keywords), the \emph{parameter keyword}
 (which specifies the attribute in the following field/s ) and \emph{flags} which simply turn on or off a given option. 

The {\verb HILLS } is a directive and switches on metadynamics. \\
The line containing  this keyword also sets up the parameters for Gaussians deposition: 
{\verb HEIGHT } (parameter keyword) followed by the Gaussian height in the energy unit of the program chosen and 
{\verb W_STRIDE } (parameter keyword), which specifies the time between the deposition of two consecutive Gaussians (in number of timesteps). 
This input has the effect of producing an additional file, called {\verb HILLS }, which has the following layout:   
\begin{verbatim}
    20.000      2.78975      0.35000      0.11111   10.000
    40.000      2.94914      0.35000      0.10926   10.000
    60.000      2.75472      0.35000      0.10737   10.000
    80.000      2.76470      0.35000      0.10542   10.000
\end{verbatim}    
This is a record of the Gaussians put during the run: it displays the time step, the center of the Gaussian (one for each CV), the width (one for each CV) and the height.
In case of well-tempered metadynamics the Gaussian height is rescaled using the bias factor printed in the last column in order to 
directly obtain the free energy (and not the bias), when summing all the Gaussians deposited during the run.   

The {\verb WELLTEMPERED } directive switches on well-tempered metadynamics. 
As explained in section \ref{methods}, CVs are sampled at a 
fictitious  higher temperature  $T+\Delta T$ determined by the bias factor $(T+\Delta T)/T$. The user must
specify this bias factor using the keyword {\verb BIASFACTOR } and the system temperature using {\verb SIMTEMP }.

The {\verb PRINT } directive allows one to monitor, during the simulation, the evolution of the CVs  between the deposition of two Gaussians.
The CVs values are printed on the {\verb COLVAR } file with a frequency, expressed in timestep units, controlled by the parameter keyword {\verb W_STRIDE }.
The file produced looks as follows: 
\begin{verbatim}
     0.000      2.26464        0.000        0.000 
    10.000      2.40452        0.000        0.000 
    20.000      2.78975        0.100        0.000 
    30.000      3.06159        0.074        0.000 
    40.000      2.94914        0.188        0.000 
    50.000      2.76442        0.185        0.000 
\end{verbatim}    
where the first column is the time step, the next contains the CV value (one for each CV), 
the third is the bias potential and the last the potential due to a wall or a restraint. 

The {\verb DISTANCE } directive selects the CV (in this case the distance between the center of mass of two groups of atoms). 
{\verb LIST } (parameter keywords) specifies the two atoms or group of atoms whose distance is calculated.
The atom indices range from 1 to  $N_{at}$  in the order they appear in a reference structure produced by the program.
In case of a group of atoms, the name of the group  must be specified between brackets {\verb <> }. 
The list of atoms belonging to the group can be placed anywhere in  the input file.  \\
The parameter keyword {\verb SIGMA } specifies the Gaussian width in CV units. 

The {\verb UWALL } directive switches on a wall potential on the collective variable  {\verb CV }. This potential
starts acting on the system when the value of the CV is greater (or lower in the case of {\verb LWALL }) then a certain
limit ({\verb LIMIT }) minus an offset ({\verb OFF }).
The functional form of this potential is the following:
\begin{equation}
V_{wall}(s)=\mathtt{KAPPA}  \left (\frac{s- \mathtt{LIMIT}+ \mathtt{OFF}}{\mathtt{EPS}} \right)^{\mathtt{EXP}}, 
\end{equation}
where {\verb KAPPA } is an energy constant in internal unit of the code,  
{\verb EPS } a rescaling factor and {\verb EXP } the exponent determining the power law.

The multiple definition of CVs is allowed.
The directive  {\verb ENDMETA  } specifies the end of the input. All the following text will be discarded.
The symbol {\verb # } is a comment line which is ignored.
 
\subsection{Umbrella Sampling}
 A general input for umbrella sampling calculation is the following:
 
 \begin{verbatim}
# a simple CV: a dihedral angle
TORSION LIST 13 15 17 1
# switching on umbrella sampling and parameters
UMBRELLA CV 1 KAPPA  200 AT -1.0 
# instruction for CVs printout
PRINT W_STRIDE 100 
# end of the input
ENDMETA
\end{verbatim}
 
 The directive {\verb UMBRELLA } switches on umbrella sampling on 
 the collective variable specified by the parameter keyword {\verb CV } (in this case the first CV that appears in the input).
 The position $s_0$ of the umbrella restraint is determined by 
 the keyword {\verb AT }, and the spring constant - whose energy units depend on the MD code used -
 by the keyword {\verb KAPPA }.
 The functional form of the potential is the following:
\begin{equation}
V_{umb}(s)=\frac{1}{2} \mathtt{KAPPA}(s- s_0)^2 \label{umbrella_potential}.
\end{equation}
\\
 The directive {\verb TORSION } selects the type of CV, in
 this case a dihedral angle defined by four atoms or group of atoms.
 \\
 The CVs value is printed on the {\verb COLVAR } file with a stride
 fixed by the keyword {\verb W_STRIDE }. In case of umbrella
 sampling, this file looks as follows:
 \begin{verbatim}
 0.000  -1.04742   0.000   0.225  RESTRAINT 1   -1.00000 
20.000  -1.09302   0.000   0.865  RESTRAINT 1   -1.00000 
40.000  -0.84990   0.000   2.253  RESTRAINT 1   -1.00000 
60.000  -1.11383   0.000   1.296  RESTRAINT 1   -1.00000 
80.000  -1.32902   0.000  10.825  RESTRAINT 1   -1.00000 
 \end{verbatim}
This file contains, from left to right: the time step, the CV value (one for each CV), the 
potential coming from the Gaussians,  the harmonic potential of umbrella sampling,
the CV on which the restraint acts and the position of the restraint.
The final calculation of the free energy as a function of this CV 
can be done using the weighted histogram analysis method,
 choosing one of the many possible implementations (see section \ref{umbrella_ex}).

\subsection{Thermodynamic integration and methods based on Jarzynski or Crooks relations}

PLUMED can be used to drag a system to a target value in CV space
using an harmonic potential moving at constant speed.
If the process is reversible, \emph{i.e.} for velocities tending to zero,  
the work done in the dragging corresponds to the free-energy difference
between the initial and the final states. 
In case of finite velocity, it is still possible to obtain an estimate of the free energy
from the work distribution using Jarzynski~\cite{jarzynski} or Crooks~\cite{Crooks98} relations.

A general input for a steered MD calculation is the following:
 \begin{verbatim}
# a simple CV: a dihedral angle
ANGLE LIST 13 15 17
# switching on steered MD
STEER CV 1 TO 3.0 VEL 0.5 KAPPA 500.0
# instruction for CVs printout
PRINT W_STRIDE 100 
# end of the input
ENDMETA
\end{verbatim}
The keyword {\verb STEER } activates the steering on the collective
variable specified by {\verb CV }. The target value is determined by
the parameter keyword {\verb TO }, the velocity, in unit of CV/kilostep,
by {\verb VEL } and the spring constant by {\verb KAPPA }.
The functional form of the dragging potential is the same as the one of 
formula \ref{umbrella_potential}.
\\
The directive {\verb ANGLE } selects the type of CV, in
 this case an angle defined by three atoms or group of atoms.
\\
The printout on {\verb COLVAR } file is analogous to umbrella
sampling with the difference that in this method the restraint position
changes during the run.

\subsection{Replica--exchange metadynamics}
When combined with GROMACS (both version 3.3 and 4), PLUMED can perform replica--exchange 
simulations coupled with metadynamics in two different ways: 
parallel tempering metadynamics (PTMetaD)~\cite{bussi_xc,camilloni_protG} and 
bias-exchange metadynamics (BE-META)~\cite{piana}.

PTMetaD is particularly useful to increase the diffusion of the system in conformational space. 
It consists in defining several replicas of the system, controlled by the same CVs but coupled with thermal baths at different temperatures $T_i$. 
As in standard parallel tempering \cite{hans97cpl,sugi-okam99cpl}, pairs of replicas can exchange at a given time $t$
two conformations $\bm{r}_i$ and $\bm{r}_j$. The probability of such an exchange is given by: 
\begin{align}
P_{i,j}=\min(1, \exp&[(\beta_j-\beta_i)(U(\bm{r}_j)-U(\bm{r}_i))+\beta_i(V_i(\bm{s}(\bm{r}_i),t)-V_i(\bm{s}(\bm{r}_j),t)) \nonumber \\
                          &+\beta_j(V_j(\bm{s}(\bm{r}_j),t)-V_j(\bm{s}(\bm{r}_i),t))]),
\end{align}
where $\beta_i=1/K_BT_i$ is the inverse temperature, $U(\bm{r})$ the internal potential, $V_i$ and $V_j$ the biasing
potentials deposited by the two replicas.  
The effect of this algorithm is to sample the degrees of freedom perpendicular to the CVs more efficiently  with respect
to standard metadynamics.

BE-META is designed to sample the system making use of a large number of CVs without the need of filling with Gaussians a high--dimensional space \cite{piana}. 
This is done employing several replicas of the system,  controlled by a few different CVs for each replica. 
Usually one defines also  a "neutral" replica, which evolves according to standard MD, \emph{i.e.} without metadynamics. 
The temperature of the system is the same for all replicas. 

The exchange probability for a pair of replicas $i$ and $j$ is: 
\begin{equation}
P_{i,j}=\min(1, \exp(\beta[V_i(\bm{s}(\bm{r}_i),t)+V_j(\bm{s}(\bm{r}_j),t)-V_i(\bm{s}(\bm{r}_j),t)-V_j(\bm{s}(\bm{r}_i),t)])).
\end{equation}

To run PTMetaD  simulations, one has to follow the standard
GROMACS procedure for parallel tempering (see GROMACS manual).  
A binary topology file must be prepared one for each replica, while
only one PLUMED input file is required. This file looks as follows:
\begin{verbatim}
# switching on metadynamics and Gaussian parameters
HILLS HEIGHT 0.1 W_STRIDE 500 
# switching on PTMetaD
PTMETAD
# instruction for CVs printout 
PRINT W_STRIDE 50 
# the CV: radius of gyration
RGYR LIST <CA> SIGMA 0.1
CA->
20 22 26 30 32
CA<-
# end of the input
ENDMETA
\end{verbatim}

The keyword {\verb PTMETAD } switches on parallel tempering plus metadynamics. 
All replicas have the same CVs, in this case the radius of gyration defined
by the group of atoms {\verb <CA> }.  
The Gaussian height set by the keyword {\verb HEIGHT } is automatically rescaled with temperature, 
following $W_i=W_0\frac{T_i}{T_0}$, where $i$ is the index of a replica and $T_i$  its temperature.
The plugin will produce one {\verb COLVAR } file and one {\verb HILLS } file for each replica.

A similar procedure is used to run BE--META.  
A PLUMED input and a binary topology file must be provided, one for each replica.  
These files must end with the replica index (\emph{e.g.}, {\verb META_INP0 }, {\verb META_INP1 }, ...) and must contain all the CVs, 
in the same order, and the keyword {\verb BIASXMD }. 
The first replica ({\verb META_INP0 }) must have the {\verb NOHILLS } {\verb CV } keyword for all the CVs; 
the other replicas must switch off the variables not used with a list of keywords {\verb NOHILLS } {\verb CV }. 
Also in this case, PLUMED will produce one {\verb COLVAR } file and one {\verb HILLS } file for each replica.

\section{Overview of the software structure}
\begin{figure}[!h]
\begin{center}
\includegraphics[height=8cm]{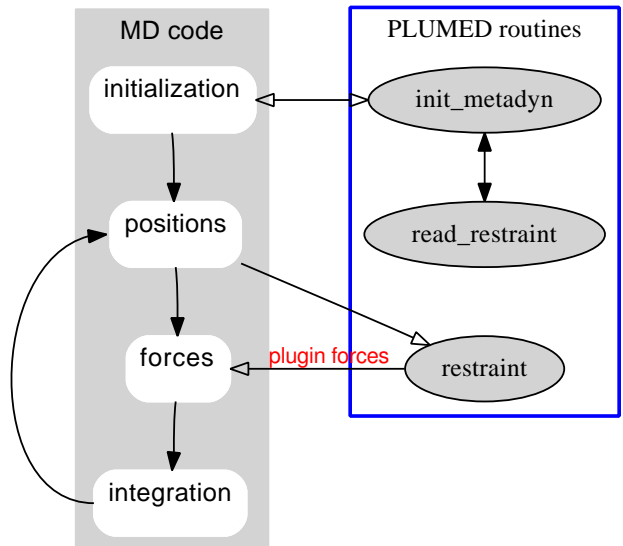}
\end{center}
\caption{Schematic representation of the interaction of PLUMED with the main MD code.}
\label{schema}
\end{figure}

PLUMED performs these basic functions:
\begin{itemize}
\item Initialization and parsing of the input file;
\item Evaluation of the CVs value for a given microscopic configuration;
\item Calculation of the forces coming from the Gaussians deposited along the CVs
          trajectory - in the case of metadynamics - or from a fixed/moving restraint acting on the CVs - in
          case of umbrella sampling/steered MD;
\item Printout of CVs value on {\verb COLVAR } file and, in the case of metadynamics, of the Gaussians deposited 
          on {\verb HILLS } file.         
\end{itemize}
The initialization of the plugin is done by the routine
{\verb init_metadyn } contained in {\verb metadyn.c }.
This routine is called by the main MD code, which communicates
to the plugin some critical information such as the number of atoms, masses
and charges, length of the simulation or timestep.
The parsing of the PLUMED input is performed by the routine {\verb read_restraint } in  {\verb read_restraint.c  }, which
reads the file and, according to the CV chosen (let's say {\verb CVname }), calls a specific parsing
routine contained in a file called {\verb restraint_CVname.c }.
\\
The second task  is controlled mainly by the routine {\verb restraint } in {\verb restraint.c },
which receives by the MD code the atoms positions at every step of dynamics. 
This routine calls a specific function, contained in {\verb restraint_CVname.c },
which calculates the CVs value and the derivatives with respect to the coordinates.
The same routine, depending on the free-energy method chosen, calls a
proper function to calculate the force acting on the atoms and
controls  the printout of CVs on the {\verb COLVAR } file. \\
The forces calculated by the plugin are communicated back to
the main MD code and added to the internal forces before the following integration
step.
The interaction of PLUMED with the principal code is summarized
schematically in Fig. \ref{schema}.
\section{Description of the individual software components}
The plugin package, distributed in a compressed tar archive, has the following directory structure:
\begin{itemize}
\item {\verb common_files }.  The directory  containing all the basic routines that compose PLUMED.
\item { \verb tests }. A variety of examples of different CVs and free-energy methods provided 
with topology and input files for GROMACS, NAMD, AMBER (SANDER module) and DL\_POLY. 
These examples, combined with a script adapted from CP2K \cite{VandeVondele:2005p10650}, 
work also as a regtest for the plugin.
\item {\verb patches }. A collection of patches to interface PLUMED with different codes  (see section \ref{instruction} for more details).
\item {\verb utilities }. Two small utilities written in Fortran: {\verb sum_hills } and {\verb driver }.
The former is a post-processing program which reads the {\verb HILLS } file produced by the plugin
in a metadynamics simulation and returns the free energy by summing the Gaussians that have been
deposited. The latter is a tool that calculates  the value of selected CVs along
a MD trajectory.
It requires a PDB file, a trajectory in DCD format and a file with the same 
syntax of the PLUMED input file.
\item {\verb doc }. A complete manual with detailed installation instructions for each code. 
\end{itemize}
 
\section{Installation instructions} \label{instruction}
The installation of PLUMED on every supported program is done through an automatic patch procedure
specific to each code. This is done on the clean code and requires its recompilation. 
All the  patching procedures are illustrated in detail in the manual.
Currently supported codes are NAMD 2.6, GROMACS 3.3.3 and 4.0.4, DLPOLY 2.16 and 2.19, SANDER (AMBER 9 version). 

\section{Test runs description}
\begin{figure}[!h]
\begin{center}
\includegraphics[height=4cm,clip]{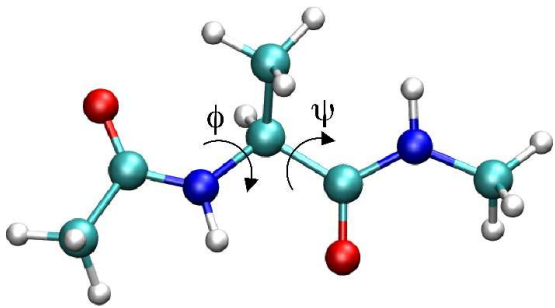}
\end{center}
\caption{Ball and stick representation of alanine dipeptide (Ace-Ala-Nme) in vacuum. The dihedral angle $\Phi$ is defined by the
set of atoms C--N--C$_{\alpha}$--C while the angle $\Psi$ by  N--C$_{\alpha}$--C--N.}
\label{diala}
\end{figure}

In the following we describe a few simple examples of the free-energy
methods implemented in PLUMED applied 
to alanine dipeptide in vacuum at 300 K (see Fig. \ref{diala}). 
The AMBER99SB force field  \cite{Hornak:2006p11531} has been used throughout all the simulations.
These test runs have been conducted with either NAMD or AMBER (SANDER module) code.
\subsection{Metadynamics}
\begin{figure}[!h]
\begin{center}
\includegraphics[height=8cm,clip]{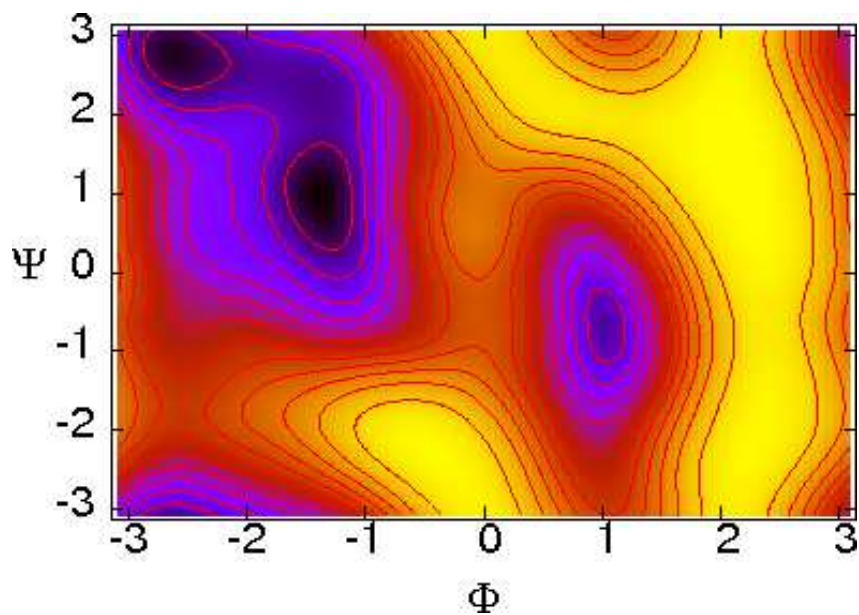}
\end{center}
\caption{Free-energy of the alanine dipeptide as a function of the two dihedral angles $\Phi$ and $\Psi$ obtained with
well-tempered metadynamics. Isoenergy lines are drawn every 1 kcal/mol.}
\label{metafes}
\end{figure}

Well-tempered metadynamics using the two dihedral angles $\Phi$ and $\Psi$ as CVs (see Fig. \ref{diala}) has been performed
with the SANDER code included in AMBER 9. The bias factor chosen is 10, 
the initial Gaussian height 0.1 kcal/mol, the width 0.35 rad for both CVs and the deposition stride 1 ps. 
The total simulation time is 5 ns.
The free energy (see Fig. \ref{metafes}) has been reconstructed from the Gaussians deposited during the run using {\verb sum_hills }, the
tool provided in the directory {\verb utilities }.
\subsection{Umbrella sampling}\label{umbrella_ex}
\begin{figure}[!h]
\begin{center}
\includegraphics[height=8cm,clip]{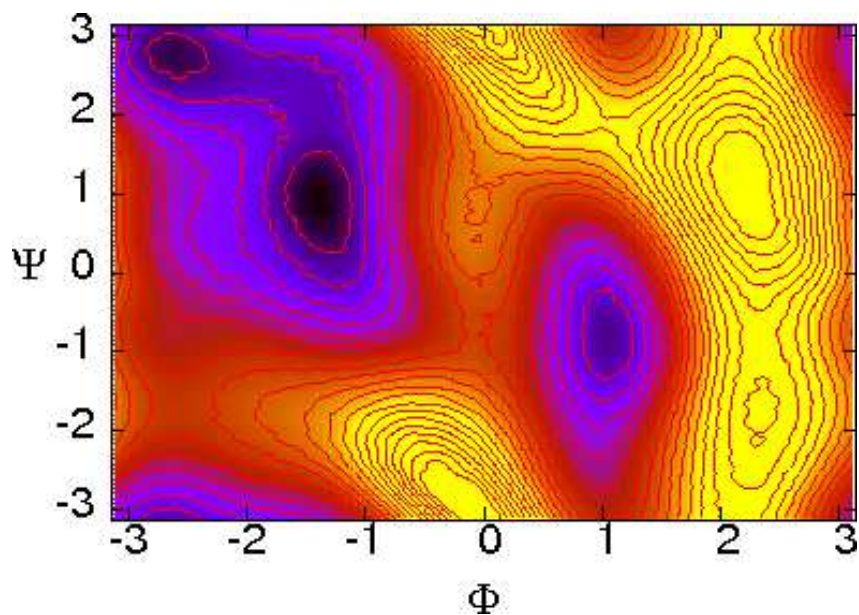}
\end{center}
\caption{Free-energy of the alanine dipeptide as a function of the two dihedral angles $\Phi$ and $\Psi$ obtained with
 umbrella sampling. Isoenergy lines are drawn every 1 kcal/mol.}
\label{umbrella}
\end{figure}

Two-dimensional umbrella sampling on the dihedral angles $\Phi$ and $\Psi$ has been performed with NAMD 
code using 676 umbrella simulations of 10 ps each and a spring constant of $100 \, kcal \, mol^{-1} \, rad^{-2}$. 
Umbrellas have been chosen in an adaptive way.
The WHAM code by Alan Grossfield \cite{grossfield} has been used to produce the free-energy profile (see Fig. \ref{umbrella}).  
\subsection{One dimensional umbrella sampling and thermodynamic integration}
\begin{figure}[!h]
\begin{center}
\includegraphics[height=8cm,clip]{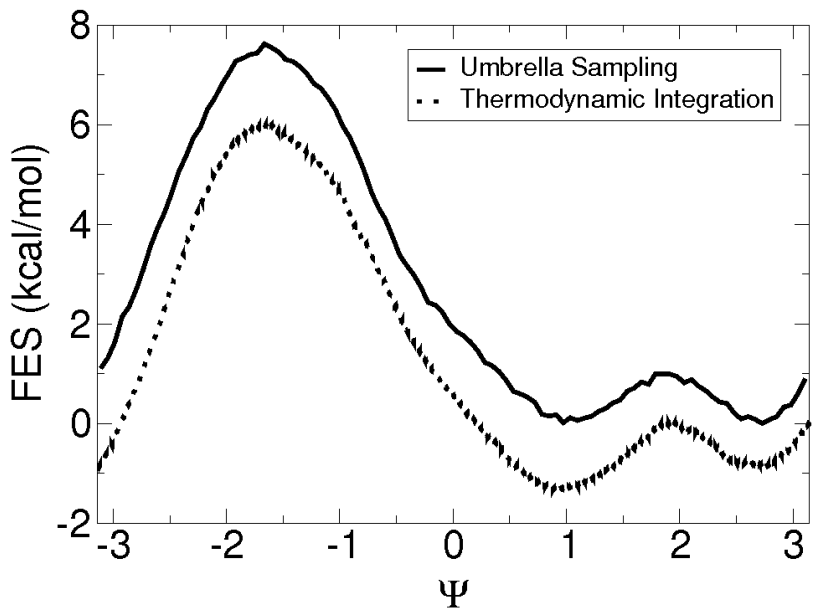}
\end{center}
\caption{Free-energy of alanine dipeptide as a function of the dihedral angle $\Psi$ obtained from a one
dimensional umbrella sampling calculation (full line) and from thermodynamic integration (dashed line).}
\label{thermovsumbrella}
\end{figure}

One dimensional umbrella sampling on dihedral $\Psi$ has been performed with NAMD code using 26 windows and running
a simulation of 20 ps per umbrella (520 ps total). The spring constant used is $100 \, kcal \, mol^{-1} \, rad^{-2}$.
Thermodynamic integration has been completed dragging the dihedral $\Psi$ from $\pi$ to  $-\pi$ in 504 ps. 
The resulting free energies are shown in Fig. \ref{thermovsumbrella}.

\section{Availability}
The plugin can be downloaded from \url{http://merlino.mi.infn.it/plumed}.
Any questions regarding the installation and usage of PLUMED can be posted to
the users mailing list at {\verb plumed-users@googlegroups.com }.

\section{Conclusions and outlook}
In this paper we have presented PLUMED, a plugin aimed at performing
the calculation of free energy landscapes using a number of
state-of-the-art methods such as umbrella sampling, steered molecular
dynamics and metadynamics. The unique feature of PLUMED is that it
can be easily ported to four popular MD codes, namely
AMBER, DL\_POLY, GROMACS and NAMD. In the next future, we plan to
further expand this list. The possibility of using PLUMED with
different host codes will allow people to choose the proper code on
the basis of its capabilities (\emph{e.g.}, implicit solvent, parallelism,
particular force fields), and also taking into account its performance
relative to a specific application.

\section*{Acknowledgements}
This work would not have been possible without the joint effort of many people in the course of the last seven years.
Among these, we should like to thank (in alphabetical order):
Alessandro Barducci, Anna Berteotti, Rosa Bulo, Matteo Ceccarelli, Michele Ceriotti, Paolo Elvati,  Antonio Fortea-Rodriguez,
  Francesco Luigi Gervasio, Alessandro Laio, Matteo Masetti, Fawzi Mohamed, Ferenc Molnar, Gabriele Petraglio and Federica Trudu. \\ 
Francesco Marini is  kindly acknowledged for his technical support,
Joost VandeVondele  for permission to use his regtest script,  
Jim Pfaendtner for giving precious suggestions in writing the manuscript.
\\
Alessandro Laio deserves a special acknowledgement for carefully reading the manuscript and giving
a number of useful suggestions.

\end{document}